# Self-Organization, Emergence, and Constraint in Complex Natural Systems

**Abstract:**

Contemporary complexity theory has been instrumental in providing novel rigorous definitions for some classic philosophical concepts, including emergence. In an attempt to provide an account of emergence that is consistent with complexity and dynamical systems theory, several authors have turned to the notion of *constraints* on state transitions. Drawing on complexity theory directly, this paper builds on those accounts, further developing the constraint-based interpretation of emergence and arguing that such accounts recover many of the features of more traditional accounts. We show that the constraint-based account of emergence also leads naturally into a meaningful definition of self-organization, another concept that has received increasing attention recently. Along the way, we distinguish between *order* and *organization,* two concepts which are frequently conflated. Finally, we consider possibilities for future research in the philosophy of complex systems, as well as applications of the distinctions made in this paper.

**Corresponding Author:**

Jonathan Lawhead, PhD
University of Southern California
Philosophy & Earth Sciences
3651 Trousdale Parkway
Zumberge Hall of Science, 223D
Los Angeles, CA 90089-0740
Lawhead@usc.edu
775.287.8005

**0. Introduction**

There's a growing body of multidisciplinary research exploring complexity theory and related ideas. This field has not yet really settled yet, and so there's plenty of terminological confusion out there. Different people use the same terms to mean different things (witness the constellation of definitions of 'complexity' itself). A good understanding of how central concepts in complexity theory fit together will help in applying those concepts to real-world social and scientific problems.

Much progress has already been made in giving an account of *emergence* in scientifically rigorous terms, and this discussion has recently gained some attention in certain corners of philosophy literature (see Collier, 2011; Hooker, 2011a, 2011b; Mossio, Bich, & Moreno, 2013; Mossio et al., 2013). While discussions of *self-organization* also abound, there is even less agreement about what it means for a system to be self-organized. Properly understood, these two terms are very closely related, and a close examination of how this is so will shed quite a bit of light on both concepts.

In **Section 1**, we will discuss emergence. After a brief overview of philosophical lineage of the concept, we will turn to a discussion of the recent advancements in complexity or dynamical systems-based reasoning that has given rise to a novel account of emergence based on facts about how systems' behaviors are *constrained*. Drawing on an example from the complexity theory literature, we will consider how the notion of "strong emergence" might be made physically meaningful through an appeal to constraints.

In **Section 2** we will expand on the notion of emergence as a constraint in the context of dynamical systems theory. We will explore the implications of multiple constraints being present in a single physical system, and think about how different constraints might interact with one another to produce complex structures. We will see that the constraint-based account of emergence manages to recover at least *some* of the intuitions associated with more traditional accounts of emergence (particularly its connection to downward causation).

In **Section 3** we will look at the concept of organization. We'll see how the language of system constraints suggests an intuitive way to understand organization in the natural world. We'll also see how a clear understanding of the physical interpretation of the mathematics underlying the constraint-based sense of emergence highlights the difference between the frequently confused concepts *order* and *organization.* Finally, we'll examine the difference between top-down and bottom-up (or "self") organization by thinking through a few illustrative examples, which provide a suggestive sketch of how the concepts articulated in this paper might apply to real-world systems.

**1.0 A Brief History of Emergence**

Philosophers have lavished a lot of words on different conceptions of emergence. The modern treatment of emergence as a concept worthy of investigation in its own right most plausibly originated with the "British Emergentists" of the late 19th and early 20th centuries (Mill 1843; Morgan 1921; Broad 1925). Like later thinking about the topic, the British Emergentists' treatment of emergence focused on a broad family of worries surrounding the relationship between the features of whole objects and the features of their constituent parts (Clayton and Davies 2006). These early discussions of emergence were centered on the then unsolved problem of life--in particular the dispute between those who posited a non-physical *elan vital* that animated living things ("vitalists") and those who saw living things as nothing more than particularly intricate (and messy) clockwork-like machines ("mechanists"). Emergence as a modern concept thus has its origins in an attempt to bring philosophy to bear on unsolved scientific problems. The elucidation of the chemical bases of life in the early and middle 20th century obviated the dispute between the vitalists and mechanists, but similar unsolved scientific problems appeared, refocusing the attention of emergentist thinkers. Discussions of emergence have found a contemporary home in the philosophy of mind literature, where mereological concerns regarding the relationship between neural cell activity and higher-level psychological features (e.g. consciousness) mirror older concerns about the nature of life.

The important point for us is that both classical and contemporary theories of emergence are grounded in an attempt to resolve certain puzzles about the natural world: emergence's history is closely intertwined with the history of science (and the philosophy of science). Emergence has traditionally been invoked when some ostensibly physical system exhibits a *novel* feature. Very roughly, a feature $P$ of a system $S$ can be said to be *novel* in this sense when (1) $P$ is distinct (in some sense[1]) from the features of the constituents of $S$ (even in aggregate) and (2) $P$ nevertheless *depends* in some sense on the features of those constituents. This asymmetric relationship between a novel feature and the features of the constituent parts of the system in which that novel feature appears has come to be called "supervenience:" one feature $P_1$ supervenes on another feature $P_2$ iff a change in $P_1$ is possible only if there is a change in $P_2$, but a change in $P_2$ is possible without a change in $P_1$. Supervenience has arisen as a key concept in many contemporary philosophical interpretations of emergence (Clayton and Davies 2006; Kim 2002; Bedau and Humphreys 2009).

## 1.1 Emergence and Causation

The supervenience relation is characterized in part by an asymmetry of causal power: while changes in the supervenient feature $P_1$ must be caused by a change in the feature $P_2$ that is supervened upon, the reverse is not true. We might think of this as an asymmetry between "upward" and "downward" causation: while the activity of low-level constituent parts can *cause* changes in higher level features, changes in those higher level features cannot *cause* changes in low-level constituents. It's clear that changes in (say) the state of the pixels on my computer screen can cause the displayed image to change from a text document to a picture, it's hard to fathom what it could even *mean* for a change in the image on the screen to *cause* a change in the state of the pixels. Any apparent "independence" of the picture from the state of the pixels is surely epistemic, born of the difficulty of deducing precisely *which* picture is being displayed based a specification of the state of pixels (and the concomitant ignorance of the precise laws governing the relationship between pixel states and displayed picture).

---

[1] This characterization is certainly vague, and reflects the widespread philosophical disagreement about what counts as "distinct enough." We shall return to this point in a moment when we consider the difference between "strong" and "weak" emergence.

This epistemic interpretation of emergence might be called *weak* emergence. This interpretation is perhaps most forcefully championed by Jaegwon Kim (Kim 1992; Kim 2000; Kim 2002), who argues that putatively emergent features can be reduced to the behavior of lower level parts through an appeal to *functions.* For Kim, an emergent feature of a system (e.g. a mental state) is nothing more than a particular functional relationship between the constituents of that system. While it may be convenient (perhaps because of epistemic limitations) to talk of tickles and pains *as if* they were novel, they are actually merely shorthand descriptions of functional relationships between more fundamental features of the system. As Mitchell (2009) argues, we might think that this account leaves emergent features bereft of causal or explanatory powers of their own and thus "stripped of any scientifically interesting features" (*ibid.* p. 32). If this Kim-style account is correct, then scientists ought to focus their investigations on the *fundamental* features of systems, not the epiphenomenal (albeit occasionally epistemically opaque) "emergent" features.

Given the burgeoning scientific interest in emergent behavior, however,[2] this leaves us with a dilemma: either the scientific community as a whole is deeply misguided in its investigation, or the Kim-style account of emergence fails to track what working scientists mean when they use the term. It is possible that the scientific definition is indeed flawed, but it is worth exploring alternative explanations. What are our other options?

Weak emergence can be contrasted with *strong* emergence, in which the asymmetrical supervenience relation between features of parts and features of wholes fails to hold. If the relationship between emergent features of a system and the features of the system's lower-level constituents is not one of supervenience, then what *is* it? It seems that we're left with two options again. On one hand, perhaps emergent features are non-physical "spooky" parts of the world (e.g. the vitalists' *elan vital*). If this is the case, however, we're faced with the same problem again: such features of systems are by definition outside the scope of scientific study. On the other hand, perhaps we're in need of a novel philosophical characterization of emergence--one that is compatible with contemporary scientific discourse. The language of constraints and boundary conditions drawn from dynamical systems theory can be leveraged to provide a better definition of the term.[3]

---

[2] Entering "emergent behavior" into Google's academic search engine yields well over 400,000 papers discussing the concept, including many which treat such features as more than epiphenomenal phantoms.

## 1.2 Emergence as a Constraint

Consider the following case drawn from Bar-Yam (2004). Suppose we've got a system of three bits that can be either on or off, and that the only allowable states of the system are those in which an odd number of bits are in the "on" state. While this is a constraint on the allowable state of the entire (i.e. 3-bit) system, it is interesting to note that it is *not* a constraint on any subset of the system, including both single bits *and* pairs of bits! Given two bits set to any arbitrary value, the value of the third bit is dictated by the global constraint. However, it isn't correct to say that any *particular* bit in the system was impacted by the global constraint, for if we were to examine *any* two-bit (or single-bit) subset, the impact of the global constraint would be totally indiscernible; given a complete three-bit state, the question "which bit's state was *caused* by the overall constraint?" makes no sense. The constraint is only apparent when we examine *ensembles* of complete three-bit states to see which are allowed and which are not. This is despite the fact that the constraint affects the state of individual bits.

But this has the air of being vaguely contradictory. On one hand, we are saying that the constraint is *globally* important only, and that the state of any one or two-bit subsystem is not affected. On the other hand, it seems obvious that the global constraint must *somehow* be affecting the value of individual bits—given two bits set arbitrarily, the constraint tells us what the third bit *must* be. So are individual bits constrained, or aren't they? Resolving this apparent contradiction requires us to shift our attention yet again—we need to attend not just to bit states or 3-bit system states, but *ensembles* of system states. Bar-Yam writes:

> The value of the individual bit is impacted by the values of the rest of the bits as far as a single state is concerned but not as far as an ensemble is concerned. This is the opposite of what one would say about the entire system, which is impacted in the ensemble picture but not in the state picture (Bar-Yam 2004, p. 20)

Notice that with or without the constraint, the set of possible states of the system is such that the probability of finding any *single* bit in a particular state is the same.

---

[3] Thanks to David Albert for suggesting this pithy way of putting the thesis.

|  | **Allowed States** |
|---|---|
| **Constrained** | {0,0,1}, {0,1,0}, {1,0,0}, {1,1,1} |
| **Unconstrained** | {0,0,0}, {0,0,1}, {0,1,0}, {0,1,1}, {1,0,0}, {1,1,1}, {1,0,1}, {1,1,0} |

In each case, each individual bit is "on" in 50% of states, and each pairing of on-off states across two bits (e.g. "first bit off, last bit on") is present in 25% of states. From the perspective of the ensemble of possible states of the system, the presence or absence of the constraint has absolutely no impact on the value of individual bits. However, if we're interested in the properties of *particular* states, the presence or absence of the constraint matters a great deal: it will dictate the allowable values of any bit in the state, given a specification of the value of the others. The ensemble statistics for particular bits are not impacted by the constraint, despite the fact that, for any given state, each bit's value is in fact constrained. On the other hand, the ensemble statistics for states of the *system* are most certainly impacted, despite the fact that the constraint operates on *bits*, rather than on system-states directly.

It's important to emphasize that while there is an epistemic aspect to this case, it is not a purely epistemic problem. The problem, in other words, is not just that we can't *predict* what the system will do given information about the allowable states of individual bits (though that's true). The problem is that we can't make that prediction even in *principle*: the global condition's impact on the value of particular bits isn't even sensibly present until we consider the behavior of bits in the context of ensembles of three-bit systems. This seems to satisfy the intuition lurking behind the traditional characterization of emergence as being a kind of "downward causation:" the state of the system *taken as a whole* plays a role in determining the allowable states of the system's constituents. The language of ensembles and constraints makes this more explicit, though, and captures the scientifically-tractable aspect of what the philosophical literature has been grappling with.

In the three-bit system, the constraint operates on each bit, and yet is not reducible to a constraint on individual bits, and does not result from the mutual influence of *pairs* of bits on one another. While this doesn't show anything about real-world systems directly, it at least suggests qualities that we might look for in our search for more concrete examples. More importantly, it demonstrates the feasibility of an account of emergence that is neither "spooky" nor diluted to the point of scientific irrelevance. Moving from the language of causes and properties to the language of constraints and systems suggests an entirely new way of thinking about emergence. Bar-Yam's case is a proof by example of the possibility that emergence might be a phenomenon that is both non-epistemic, and also scientifically meaningful: *to characterize a system's feature as "emergent" is to say something about the nature of the constraints the feature's presence imposes on the dynamical form of the system.*

## 2.0 Emergent Constraints in Dynamical Systems

Discussions of self-organization abound in the complexity theory literature (Waldrop 1992; Kauffman 1993; Auyang 1998; Strevens 2003; Gribbin 2004; Mitchell 2009; Hooker 2011a; Johnson 2009; Prokopenko, 2013), but getting a clear definition of *organization* is startlingly difficult. Most authors seem to take it for granted that we have an intuitive grasp on what it means for a system to be organized. Perhaps the most succinct definition comes from physicist Sunny Auyang, who says that organization is the "formation of new structures in the symmetry-breaking of equilibrium systems" (Auyang 1998, p. 242). Another good suggestion is given by Cliff Hooker, who writes that self-organization is "a process where dynamical form is no longer invariant across dynamical states but is rather a (mathematical) function of them" (Hooker 2011b, p. 212). Both of these are useful initial characterizations of the phenomenon, and there's a sense in which each of them captures an important feature of organization. However, it's going to take quite a bit of unpacking to figure out just what even these two characterizations are driving at, and to articulate how they relate to the kind of emergence discussed so far.

Bar-Yam's 3-bit system, discussed in **Section 1**, shows the possibility of genuine emergent constraints in a toy system. It is, however, also somewhat limited. Because the system is so simple—because it consists of states of only three bits, each of which can take on only one of two possible values—it is not obvious how to translate Bar-Yam's formal insight into an insight about the workings of real-world physical systems.

## 2.1 Patterns and Constraints

Bar-Yam's example can be represented as an abstract space of possible states of the system. Every point in the space represents a specific value for each of the three bits. This approach should be familiar: it is the notion of a *state-space* that's commonly employed in many sciences. If we had some facts about the dynamics of Bar-Yam's system—some kind of pattern that described how the states transition from one to another—then we'd be able to plot out a map of the system. If the dynamics were totally deterministic, then for any starting position in the space, we'd have a *path* through the space that represents the succession of states the system would proceed through if it started in a given state. Given a picture like this, how do we understand the kind of system-wide constraint that Bar-Yam describes? The answer should be fairly obvious: the constraint represents points (or regions) of the state-space which are, so to speak, "out of bounds"—states that the system simply *can't* get into, no matter what its dynamics are, or where it starts, at least so long as the emergent constraint remains in place.

Note that this constraint is different in kind from either initial-condition or boundary-condition constraints. The difference between an emergent constraint and an initial-condition constraint is clear (an initial condition just defines where the system *starts*, but says nothing about where it is going), but the difference between an emergent constraint and a constraint imposed by a boundary condition is a bit less obvious. The difference is most striking if we think about the sort of state-space described—the one representing Bar-Yam's three-bit system—as being embedded in a larger state space representing a system of $n \ggg 3$ bits. If we were to define some dynamics for the system, a boundary condition would define a topologically connected subspace (or, in the case of the specific example at hand, a sub-*graph*) to which we should restrict our attention. The only restriction a boundary condition places on perturbations of some system state is that those perturbations cannot take the system *as a whole* outside the subspace—i.e. into regions of the space where more than three bits have possible values. It has nothing whatsoever to say about transitions of individual bits within that space. Contrast that with the emergent parity constraint in Bar-Yam's case: in addition to restricting states of the system as a *whole* the emergent constraint (as we saw) also plays an important role in determining the state of *individual* bits—at least when they appear in context. We can see this even more clearly when we ask how much information we need to specify the successor-state to some given system-state. Given a state of three bits, we can ask "if I were to flip one bit at random, what's the probability that the resulting state will be one that is permitted by the constraints operating on the system?" The kind of boundary condition we just suggested would have nothing at all to say about this question, while the emergent parity constraint would definitely play a role in our calculation. While both traditional boundary conditions and emergent constraints restrict the time-evolution of the system in various ways, they are distinct concepts with distinct physical interpretations.

What's going on here? Notice that in adding a constraint like the emergent one, we're increasing the *pattern richness* of the space of possible states into which the system can transition. If multiple constraints are operative on the same system at the same time, they'll have to be mutually-consistent if the system is to be sensibly thought of as a single entity. Consider, for example, the stipulation that in addition to the first constraint given on Bar-Yam's parity system, we add the constraint "the sum of the values of all of the bits must be one." Clearly, this additional restriction constrains the available states of the system even further—now only <1,0,0>, <0,1,0>, and <0,0,1> are legal. On the other hand, adding the constraint "the number of 'on' bits must be even" is (manifestly) not allowed, as it's being in place is *ruled out* by the original emergent constraint given by Bar-Yam. There's just no way for the system to get into a state where both of those constraints are satisfied. Multiple restrictions placed on the same space restrict the possible states that the system can get in to. What is perhaps less obvious, though, is that the addition of an emergent constraint on a system *also restricts what other emergent constraints can be applied to the same state in the future*. Why is this significant?

For most interesting systems, the choice of which state-space to use is partially a matter of our predictive and explanatory goals, as there is more than one reasonable choice (McAllister, 2003; Lawhead 2014). Consider the system corresponding to the kitchen in my apartment. We have a variety of choices in what kind of state-space to use when we describe this system. The one based on Newtonian Mechanics—the *phase space*—seems like a safe enough choice; spaces like that have been pretty well-mapped, and the maps are very well-confirmed by experiment. Given that choice, consider what it means to say that the space is "well-mapped:" it means we know a tremendous amount about the *dynamics* of the system. In this case, it means that the patterns that underlie the motion of points inside the state-space corresponding to Newtonian Mechanics are fairly well-understood. Next, consider what it means to say that Newtonian mechanics *applies* to my apartment in the first place. As we said above, a set of dynamics for a given state space provides a set of *directions* for moving from any point in the phase space to any other point—it provides a *map* identifying where in the space a system whose state is represented by some point at $t_0$ will end up at a later time $t_1$. This map is interesting largely in virtue of being valid for any point in the space: no matter where the system starts (as long as it starts somewhere in the space) at $t_0$, the dynamics will describe a set of *patterns* in how its state changes. That is, given a list of points $[a_0, b_0, c_0, d_0...z_0]$, the dynamics give us a corresponding list of points $[a_1, b_1, c_1, d_1...z_1]$ that the system will occupy after a given time interval has passed (assuming that in the interim, the system's path didn't take it *outside* the space; we'll discuss this point shortly).

As we said, though, this is not the only possible approach. In addition to the possibility of choosing a different kind of state-space with which to describe my apartment (if we're masochists, maybe something like a Fock space in quantum field theory), it might be the case that there are also patterns to be discerned in how certain *regions* of our chosen space— the Newtonian phase space, in this case—evolve over time. That is, we might be able to describe patterns of the following sort: if the room starts off in any point in region $P_0$, it will, after a given interval of time, end up in another region $P_1$. This is, in fact, the form of the statistical-mechanical explanation for the Second Law of Thermodynamics. This is clearly not a description of a pattern that applies to the space in general: there might be a very large number (perhaps even a *continuous infinity* if the space in question is continuous) of points that do not lie inside $P_0$, and for which the pattern just described just has *nothing to say*.

What does all this have to do with emergence and organization? Well, consider what it means to say that for some system *S*, there are a plurality of different state spaces we might choose from to represent it. For a normal, living human brain, for instance, we might choose a space defined in neuroscientific terms (in which points in the space represent action potentials, neurotransmitter location, &c.), or we might choose the space defined in terms of organic chemistry (in which points in the space represent the positions and properties of chemical molecules, &c.), or we might choose the space defined by good old Newtonian mechanics (which we're already familiar with), and so on. We might even choose the space defined by *cognitive neuroscience*, in which points in the space represent information-processing states of functional collections of brain regions.

The multiplicity of interesting (and useful) ways to represent the same system—the fact that *precisely the same physical system* can be represented in very different state spaces, and that interesting patterns about the time-evolution of that system can be found in each of those state spaces—has tremendous implications. Each of these patterns, of course, represents a *constraint* on the behavior of the system in question; if some system's state is evolving in a way that is described by some pattern, then (by definition) its future states are *constrained* by that pattern. As long as the pattern continues to describe the time-evolution of the system, then states that it can transition into are limited by the presence of the constraints that constitute the pattern. To put the point another way: *patterns in the time-evolution of systems just are constraints on the system's evolution over time*.

It's worth emphasizing that these constraints can (and to some degree *must*) apply to all the spaces in which a particular system can be represented. After all, the choice of a state space in which to represent a system is just a choice of how to *describe* that system, and so to notice that a system's behavior is constrained in one space is just to notice that the system's behavior is constrained *period*. Of course, it's not always the case that the introduction of a new constraint at a particular level will result in a new *relevant* constraint in every other space in which the system can be described. For a basic example, visualize the following scenario.

### 2.1.2 An Analogy Using Construction Paper

Suppose we have three parallel Euclidean planes stacked on top of one another, with a rigid rod passing through the three planes perpendicularly (think of three sheets of printer paper stacked, with a pencil poking through the middle of them). If we move the rod along the axis that's parallel to the planes, we can think of this as representing a toy multi-level system: the rod represents the system's state; the planes represent the different state-spaces we could use to describe the system's position (i.e. by specifying its location along each plane). Of course, if the paper is intact, we'd rip the sheets as we dragged the pencil around. Suppose, then, that the rod can only move in areas of each plane that have some special property—suppose that we cut different shapes into each of the sheets of paper, and mandate that the pencil isn't allowed to tear any of the sheets. The presence of the cut-out sections on each sheet represents the constraints based on the patterns present on the system's time-evolution in each state-space: the pencil is only allowed in areas where the cut-outs in all three sheets overlap.

Suppose the cut-outs look like this. On the top sheet, almost all of the area is cut away, except for a very small circle near the bottom of the plane. On the middle sheet, the paper is cut away in a shape that looks vaguely like a narrow sine-wave graph extending from one end to another. On the bottom sheet, a large star-shape has been cut out from the middle of the sheet. Which of these is the most restrictive? For most cases, it's clear that the sine-wave shape is: if the pencil has to move in such a way that it follows the shape of the sine-wave on the middle sheet, there are vast swaths of area in the other two sheets that it just can't access, no matter whether there's a cut-out there or not. In fact, just specifying the shape of the cut-outs on *two* of the three sheets (say, the top and the middle) is sometimes enough to tell us that the restrictions placed on the motion of the pencil by the third sheet will likely be relatively unimportant—the constraints placed on the motion of the pencil by the sine-wave sheet are quite stringent, and those placed on the pencil by the star-shape sheet are (by comparison) quite lax. There are comparatively few ways to craft constraints on the bottom sheet, then, which would result in the middle sheet's constraints *dominating* here: most cutouts will be *more* restrictive than the top sheet and *less* restrictive than the middle sheet[4]

The lesson here is that while the state of any given system at a particular time has to be *consistent* with all applicable constraints (even those resulting from patterns in the state-spaces representing the system at very different levels of analysis), it's not quite right to say that the introduction of a new constraint will *always* affect constraints acting on the system in *all* other applicable state spaces. Rather, we should just say that every constraint needs to be taken into account when we're analyzing the behavior of a system; depending on what collection of constraints apply (and what the system is doing), some may be more relevant than others.

The fact that some systems exhibit interesting patterns at many different levels of analysis—in many different state-spaces—means that some systems operate under far more constraints than others, and that the introduction of the right kind of new constraint can have an effect on the system's behavior on many different levels. The lesson to take from our discussion in **Section 1** about Bar-Yam's toy system is that *"emergence" just means the introduction of a new instance of a particular kind of constraint on allowable states of the system: a constraint that significantly alters the dynamical form of the system at multiple levels of analysis.* A feature of a system is emergent just if its appearance significantly alters the dynamical form of the system across disparate levels of analysis. This explains why emergent phenomena seem to exhibit features that have been traditionally associated with "downward causation;" they are restricting the allowable states of the system, and this restriction can manifest in changes to the dynamical form—the patterns in its state-transition—of the system at multiple levels of analysis. Unless we appreciate the relationship between the patterns at different levels, this can look incredibly mysterious—even anomalous—as the factors constraining a systems' behavior from one perspective might not be apparent from another perspective (in virtue of changes in how the system is represented). Once we see that any system's behavior must be consistent with *all* the patterns that describe its behavior—and that in order to see some of those patterns, we must shift our perspective, as we did when we started paying attention to *ensembles* of states rather than single states (or single bits) in Bar-Yam's system—the mystery starts to dissolve. Emergent constraints only seem more mysterious than more mundane constraints (like boundary conditions) in virtue of the fact that their impact is spread out across multiple levels of analysis.

---

[4] Deacon (2012) discusses emergence and constraint, but is marred by this confusion as he suggests that constraints in the sense of interest to us here *just are* boundary conditions under which the system operates.

### 3.0 From Emergence to Organization

The observation that emergence is can be understood in terms of mutually-interacting constraints operating at highly disparate scales and levels of analysis is not novel.[5] However, the preceding discussion has demonstrated that this perspective is perhaps not as at odds with more traditional accounts of emergence as it may first appear to be. In particular, the dynamical account of mutually interacting constraints operating at disparate scales preserves many of our intuitions about the role of downward causation and supervenience in discussions of emergence. This should be seen as a virtue of this view. Let us now begin to expand the constraint-based account beyond emergence, and see how it leads into a definition of self-organization. But first, a brief aside is in order.

### 3.1 Aside: Order vs. Organization

One more issue is worth flagging here before we begin to consider the relationship between emergence and self-organization: the distinction between organization and *order*. The conflation of these concepts is so widespread as to be nearly ubiquitous. Auyang refers to self-organization as "the spontaneous appearance of order, which is common in complex systems," (Auyang [1998] p. 32), but she's far from the only guilty party, and the mistake cuts across wide disciplinary boundaries. Kelly (2010) includes a lengthy discussion of emergence and complexity-increasing processes in biological and technological systems, but consistently slides back and forth between calling these processes "order-increasing" and "organization-increasing." Stuart Kaufmann is even guilty of the mistake in his seminal 1993 book, writing, "Simple and complex systems can exhibit powerful *self-organization*. *Such spontaneous order* is available to natural selection and random drift for the further selective crafting of well-wrought designs or the stumbling fortuity of historical accident (Kaufmann 1993, p. 1, emphasis mine). Kaufmann uses the terms 'spontaneous order' and 'self-organization' as if they were synonymous (or very nearly so), and the conflation has largely passed without remark, with very few exceptions.[6]

---

[5] The implications for (in particular) complex adaptive *biological* systems has been well-explored. See (Barandiaran & Moreno, 2006; Collier, 2011; Cumming & Collier, 2005; Mossio, Bich, & Moreno, 2013; Mossio, Saborido, & Moreno, 2009; Mossio & Moreno, 2010; Saborido, Mossio, & Moreno, 2011).

We've now assembled all the tools we need to articulate the distinction between organization and order, and to discuss how emergence bears on that distinction; most of the heavy lifting has already been done in the preceding pages. The existence of a real pattern in one of a system's state-spaces can represent a restriction on the movement of the same system in other of its state-spaces (though not all constraints must do this, as the addition of a new constraint might only restrict the system from evolving into states that were *already* forbidden by other existing constraints). The more patterns that exist in a system, the narrower the field of possible states that the system can transition to— and so the more convoluted its dynamics become. If emergent constraints are those which have an impact on the dynamical form of a system at multiple levels of analysis, then organization is just the process by which emergent constraints actually *emerge*. This is what Hooker was getting at when he said that organization is "a process where dynamical form is no longer invariant across dynamical states but is rather a (mathematical) function of them" (Hooker 2011, p. 212). Emergent constraints are constraints not just on a system's *state*, but also on its *dynamics:* rather than just restricting which points a system can occupy in its state-space, they restrict how the system is allowed to *transition* from one point to another. Similarly, rather than depending just on which point a system occupies in its state-space, emergent constraints depend on its dynamical history.

This exposes why 'organization' cannot possibly be the same thing as 'order' in any traditional sense of the term. A system that is highly ordered is, in some sense, also a system with quite a stable dynamical structure. Crystals, for instance, are highly ordered structures in virtue of being spatially

---

symmetric, low-entropy, and relatively *static* systems (*ibid)*. They are quite stable, but only in virtue of lacking a large number of interesting patterns that describe their time-evolution. Crystals have the same response to a fairly wide class of environmental perturbations: just sit there (and maybe resonate a little bit), and this characteristic behavior is a result of the fact that crystals have a lattice-like atomic structure in which the force exerted on each atom by each of its neighbors is uniform and constant.

By contrast, highly *organized* systems tend to be very *interesting*, and *dynamic* systems: systems which (to borrow Stuart Kaufmann's evocative turn of phrase) live forever on the edge of chaos. To exhibit the kind of multi-layered diversity of patterns that characterizes organized systems, significant symmetry-breaking of some sort is virtually always required, and functional differentiation is an easy (and common) route to increased organization. This is likely what led Auyang to characterize organization in general as the "formation of new structures in the symmetry-breaking of equilibrium systems" (Auyang 1998, p. 242). The initial conflation of order and organization likely stems from the fact that both highly ordered (low-entropy) systems and highly organized systems occupy positions in their state spaces that are, in some sense, "special." A highly ordered system is one with very low entropy—one that is in a microstate corresponding to a low-volume macrostate. A highly organized system's location in state space is also unusual, but it is unusual in a very different sense: rather than corresponding to a very low-volume macrostate, it is a location that is rich in patterns at many different levels of analysis and which features many patterns that impact the system's dynamical form across multiple levels. A highly organized system might *also* be a low-entropy system (and dissipative systems will have to pay for their increased organization through an increase in environmental entropy, just as they would with any other sequence of state-transitions), but a system that is low-entropy *and* highly organized is special in two very different senses. *Order* is a feature of a *state* in which a system might find itself, and this is relatively independent of the dynamics of that system. Organization, on the other hand, is inextricably linked to the dynamical structure of systems, as it describes constraints on how systems can change. These two properties of physical systems should be thought of as orthogonal to one another: a system may be highly ordered but lack much organization--for instance a very low-entropy quantity of homogeneous hydrogen gas--highly organized but rather disordered--for instance a caste-differentiated ant colony that has recently been disturbed by a backhoe (Gordon 2010)--or any other combination of the two. The processes by which *order* appears in natural systems are distinct from the processes by which *organization* appears, and treating the two as interchangable is to invite significant confusion.

### 3.2 Self-Organization and Top-Down Organization

Let us now turn to the problem of distinguishing self-organization from other instances of organization. There are (at least) two distinct ways in which increased organization in a system could come about: it could be imposed from "outside" the system, or it could result from the dynamics of the system itself without significant external influence. The key attribute here is a certain kind of *dynamical symmetry* between the ways in which a system shapes the environment in which it is embedded and the ways in which that environment shapes the system. Attending to imbalances in the dynamical influence between the organizing system and the organized system can give us a natural set of criteria to help delineate self-organization from the other sort of organization, which we might call *top-down* organization, for reasons that should become apparent soon.

Let's proceed by considering a few hypothetical cases and seeing how our intuitions accord (or fail to accord) with more formal concepts in complex systems theory. Consider the following two lines of development: the organization into functional groups that the human brain undergoes between birth and adulthood, and the organization into similar functional groups imposed on a lab-designed copy of that brain. Neural networks learning to navigate a complicated environment are paradigmatically self-organizing systems, but in virtue of what is that true? Organization is properly understood as involving two processes: an increase in pattern-richness in the dynamical equations describing the time-evolution of a particular system, and a *decrease* in the total volume of the state-space regions into which the system can find its way[7].

The human brain is (of course) a neural network composed of nodes (in the form of neurons, glial cells, &c.) and edges (in the form of of axons, dendrites, &c.). The sheer number of nodes at birth is significantly greater than the number of nodes will be at any other point in the lifetime of the neural network: most humans are born with something like one trillion neural cells, while most adults have had that number reduced by an order of magnitude (to ~200 billion neural cells). However, this reduction in *number* of nodes is accompanied by an *increase* in the information-processing capacity of the network as a whole. At first, this fact looks somewhat puzzling, but a clear understanding of organization shows that this is precisely what we ought to expect, given *how* the brain goes about pruning away nodes. As the brain develops, it becomes more organized in the sense given above. The one trillion neural cells constituting the infant's neural network are sparsely connected (especially when compared to the structure of the network in a mature adult), and the network as a whole exhibits a tremendous lack of differentiation; functional groups have yet to strongly emerge as specialized regions dedicated to particular tasks, and the network topology of the brain is highly symmetric, resembling a crystal more than the highly-organized (and far from equilibrium) network it will eventually become.

---

[7] That is, a decrease in the volume of the state-space regions that the system can find its way into *without* a significant change in dynamical form—without a loss of organization by way of a catastrophic phase change.

As the infant interacts with an active ambient environment, its neural network adapts to better solve the information-processing tasks it encounters on a regular basis. The slogan for this phase of neural development, oft-repeated in undergraduate cognitive neuroscience textbooks, is telling: "neurons that fire together wire together." Crucially (at least for our purposes), the progress from undifferentiated infant neural network to specialized (and highly differentiated) adult neural network involves two primary changes: the construction of new edges between nodes on the network—neurons "wiring together,"—*as well as* the pruning of nodes that fail to play important roles in the construction of functional structures designed to solve whatever problems the neural network regularly encounters in the course of its development.

Consider this process in light of what we've said about organization's relationship to state-constraint and multi-level patterns. The undifferentiated trillion-plus neural cells in the infant's brain suggest a tremendously large neural state-space for the infant's neural network. As the infant moves toward adulthood, though, the state-space transforms along with the network itself. As neural cells die off, the total size of the state space (considered from a neurobiological perspective) shrinks accordingly—this is true in just the same way that removing a number of molecules of gas from a container will reduce the dimensionality of the thermodynamic phase-space associated with the container. With fewer constituent parts composing the system, the dimensionality of the state space gradually shrinks. By the time our intrepid infant has reached young adulthood (having lost billions of neural cells along the way, undergraduate party attendance or no), a number of possible neuronal states which *might* have been accessible to his neural network at birth are simply closed off, as there are no longer enough individual cells to make them possible.

This point, I take it, is relatively obvious. Perhaps less obviously, the *emergence* of differentiated functional groups—which, as the infant grows toward adulthood, take on increasingly specialized roles in processing specific classes of information about the world around her—also constrains the range of possible states into which the neural network might evolve. These constraints, however, are not the result of decreased dimensionality of the state space as a whole, but rather a consequence of the explosion of many new different (but mutually-consistent) patterns in the time-evolution of the states of the neural network. The fact that the joint activation of two or more nodes increases the strength of the connection between those nodes is instrumental in shaping the *dynamical* form of the neural network. In contrast, the morphological form of the network is shaped primarily by the pruning operations that remove under-utilized nodes and edges. One consequence of this increasing *organization* of the brain into functional groups is that some neuronal states—those that are frequently needed to discharge important functions—become quite easily accessible from a wide variety of prior states, while other neuronal states—those which would require patterns of activation that have not been reinforced (or have even been discouraged through negative feedback to the network)—become increasingly difficult to access from virtually all prior states. The patterns in the time-evolution of the functional (or psychological) states constrain the patterns in the time-evolution of neurons.

For a mundane example of this phenomenon, consider the difficulty of producing certain classes of motor impulses which conflict with long-rehearsed ways of moving (most children will encounter this sort of puzzle first in the form of an adult challenging them to "rub your belly and pat your head at the same time"). The more practiced the rehearsed motion (and the more the newly-attempted motion interferes with the well-rehearsed pattern of neuronal firing), the more difficult it is to actually produce the desired neural activity. Consider also the sort of "blocking" you might experience when trying to learn Italian if you already know Spanish; the similarity of the two languages can actually make acquiring vocabulary and grammar that's just *slightly* different more difficult than it otherwise would be. Even more strongly, some general patterns of neural activation simply *exclude* other patterns for as long as they are active; for as long as my brain maintains the sequence of state-transitions necessary to keep me awake, sober, and conscious, enormous volumes of my neural state space are simply unavailable to me (e.g. paths through the state space which correspond to each neuron in my brain firing in rapid succession in spatial order from the top left anterior corner of my brain and working their way down, backward, and to the right). The organizational patterns that have developed over the course of my life, in other words, simply prohibit my brain's transition into a vast number of different possible locations in its state space. Biofeedback or cognitive behavioral therapy (CBT) based treatments for some psychological conditions (like panic disorders) capitalize on this capability, and teach patients to consciously impose certain top-down constraints on their neurological activity. This is a hallmark of organization: the *emergence* of mutually-consistent patterns in the time-evolution of a system which, so long as they persist, "block off" a number of other possible states and constrain other possible time-evolution patterns in the system.

Note that the particular *sort* of organization that a given neural network undergoes is very heavily influenced by the structure of the ambient environment in which the network develops. While it is surely true that the organizational process relies heavily on particular features of the environment, it seems to me that the most important feature for us to attend to is not *whether* two systems interact, but rather the *nature* of the interaction. Human neural networks rely on very particular kinds of interactions with very particular kinds of environments in order to organize in anything resembling a "normal" way. Indeed, some cognitive scientists have argued that our reliance on the environment goes deeper than merely shaping our early cognitive development, suggesting that the specific kind of organization that most human neural networks evince actually depends on *ongoing* coupling with environmental props in order to get the most out of its organizational scheme (Clark 2003; Clark 2002; Clark and Chalmers 1997; Adams and Aizawa 2011).

However, it's also important to emphasize that this influence goes the other way too: the developing brain has an increasingly significant amount of control over its environment. As a child's neural network becomes more highly organized (and better at information processing), the role the child plays in shaping her own environment—and thus in steering the network's own continued organization—becomes increasingly important. As a child develops interests and preferences (along with the cognitive and bodily resources to manipulate the world), she begins to influence the world around her by seeking out things that interest her, avoiding things that bore her, associating with people who provide the sort of stimulation she prefers, &c. The dynamics of the neural network in this standard case become more and more relevant for determining the future organizational pressures on the network itself, and this kind of give-and-take with the environment becomes only more pronounced as the network becomes more highly organized. This kind of dynamical symmetry—where an organizing system is shaped by its environment, but also becoming an increasingly dominant force in shaping that environment—is the right way to understand *self*-organization.

Consider the difference between the processes that produce an organized human neural network and the production of the same network through different means. Suppose, for instance, that we could construct a similar (or even identical) neural network by very subtle manipulation of matter at the atomic level. If this copy is a good one, then surely it is just as organized as the "original" from which it was copied. Still, it seems relevant that the new neural network was constructed from the "top down," in accord with some kind of master plan. In the standard case, recall, the neural network from which this atom-for-atom copy might be produced is shaped by a myriad of environmental pressures, and organizes as a result of its interplay with an active environment. In contrast, the "vat-grown" copy of the same brain is produced by environmental interactions that are far more one-sided: the machinery constructing the copy of the brain must painstakingly put each atom into place, and thus the brain-copy must in *some* sense interact with an active external environment. The point of departure, though, is in asymmetry between the environment's influence on copy-brain and copy-brain's influence on the environment.

In the standard developmental case, a child's neural network becomes organized as a result of interactions with an ambient environment that the child herself plays a significant (and increasing) role in shaping. On the other hand, the kind of influence that the dynamics of the copy-brain grown in the lab have on the ambient environment aren't particularly important in determining the course of copy-brain's development. Copy-brain develops as a result of pressures that are put on its formation by the machines (and scientists) responsible for putting each atom in its proper place, and its eventual structure corresponds to some sort of master plan in the mind of a scientist (or a blueprint on a hard drive somewhere). Despite ending up similarly organized, the paths that copy-brain and a standard child's brain take to get to that level of organization are tremendously different, and it seems that this difference effectively captures the crux of self-organization: copy-brain's process of organization has far less of an impact on its environment than does the process of producing similar organization by more standard means.

### 3.3 Self-Organization, Autonomy, and Control

To put the point succinctly (if rather loosely), *a self-organizing system is one which is among its own biggest influences*. This requires some qualification, as it certainly isn't correct to say that (for instance) an infant's brain wouldn't end up behaving significantly differently if its developmental environment had been absent. Quite the opposite is true: self-organized systems are malleable, but are malleable in a very particular way: they're subject to environmental influences, but also play an important role in shaping their environment. Hooker's earlier characterization (quoted above) of organization gets this right: a self-organized system's dynamical form is not time-independent, and its increased organization consists in part of losing any time-independence it might have once had. However, the preceding discussion helps us go beyond Hooker's definition. While both self-organized and top-down organized systems evince dynamical structures that are not straightforwardly functions of time, a *self-organized* system's dynamical structure is a function of time, environmental inputs, *and* its past and present dynamical form. Moreover, as self-organized systems gain more autonomy, the last part of this function increasingly dominates, and the system increasingly becomes its own most significant influence.

Perhaps a more precise way to put the slogan of the last paragraph, then, is to say that a system is self-organized when some or all of its emergent constraints are the result of feedback loops operating within the system itself, rather than between the system and its ambient environment. A self-organizing system is one which has some significant degree of control over its own emergent constraints. This partially explains the close connection between self-organization and our intuitions about complexity: complex physical systems (very roughly) are those which have many components interacting with one another in non-trivial ways across many different scales (Cumming & Collier 2005; Lawhead 2014; McAllister 2003; Mossio et al. 2013; Ryan 2007). In cases of highly organized stable complex systems, self-organization is the most likely mechanism by which this sort of structure might appear in the absence of design.

## 4. Conclusion

The arguments I've advanced here have the potential to help shed light on a number of different characteristics that complex systems tend to share with one another. To close, I will mention a few of these connections, which might be taken up in investigating the foundations of complexity theory. Organized complex systems often show behavior that is highly path-dependent; Murray Gell-Mann famously described complex systems as being the result of the accumulation of frozen accidents (Gell-Mann 1995). This characterization of self-organization helps expose the mechanism by which such accidents continue to matter for self-organized complex systems long after they've been frozen. Because self-organizing systems are among their own biggest influences, tiny differences between two otherwise similar self-organizing systems can, given the right conditions, nudge the two systems in very different developmental directions and reinforce their initially small differences. The formation of preferences during the development of a normal human's brain is an example of this phenomenon: early exposure to (for instance) one flavor of ice cream rather than another—Gell-Mann's frozen accident at its tastiest—can result in a self-reinforcing preference for a particular flavor profile later in life, playing a role in shaping a person's gustatory habits for years. Less positively, educators are familiar with the phenomenon of "learned helplessness" for particular subjects, in which bad early-life experiences with a particular discipline (mathematics is a common one) becomes a self-reinforcing difficulty with the subject matter: the first few bad experiences lead the student to avoid the discipline, which makes it harder to do well, which leads to further bad experiences. Self-organized systems are vulnerable to problems like these in a way that outside-organized systems are not, and how to incorporate that fact into the design and manipulation of those systems is a question worth exploring.

In addition, the characterization of organization in general I've given here suggests an intriguing relationship between organization, autonomy, and adaptation. The multiple inter-influencing patterns present in the time-evolution of highly organized systems make it possible for them to respond to a diverse class of environmental perturbations with different behaviors: organization creates opportunities for adaptation by increasing the range of dynamically interesting responses to an ambient environment. At the same time, the presence of the constraints on the time-evolution of organized systems prevents them from responding strongly to just *any* environmental input; an organized system can match its range of possible actions to an active environment, and can work to maintain favorable conditions in its local environment. An organized system can be highly sensitive, capable of nuanced responses to small changes in the world around it, but (in virtue of the variety of constraints operating on it) only sensitive to the right *kinds* of inputs. The right combination of sensitivity and restrictions might well lead to increasingly nuanced responses to the environment, which could create the basis for an adaptive edge. On the other hand, highly organized systems are likely to be more vulnerable to certain kinds of damage too, as external influences that force the system into a state that is not compatible with one (or more) of its emergent constraints might well have disastrous consequences for the system; with functional differentiation comes a degree of fragility. Organization might, then, be seen as a trade-off between flexibility and stability. This suggests an important question : just what kinds of environmental pressures create selective pressure toward increased organization? What is the relationship between those pressures and the development of intelligence? There is an intriguing potential to contribute to the explanation of why certain aspects of biological evolutionary theory seem to describe the development of so many non-biological systems (see, e.g., Kelly 2010; Zurek 2004). If what we have said here is correct, a careful study of organization-increasing processes has the potential to shed light on the formation of intelligent, self-regulating systems of both the biological and nonbiological variety.

On a more practical note, the study of self-organization is likely to become increasingly important as humans begin to not only attempt to *understand* self-organized systems, but also to attempt to *engineer* or *design* them (Prokopenko 2013; Prokopenko 2009). Most urgently, the question of whether (and how) to approach the task of *geoengineering* a solution to the looming global climate crisis is just over the horizon. To the extent that the global climate is a self-organized complex system, attempts to deliberately engineer the future of the climate by intervening in the behavior of the system now amount to clear attempts at guided self-organization. Without a solid theoretical understanding of self-organized systems--including how they're different from more traditionally organized systems--this undertaking is even more fraught with risk.

Finally, there remains the task of integrating this account of emergence into a more general theory of scientific laws. If emergence is a phenomenon of scientific interest *and* does not conflict with the spirit of reductive physicalism, then how are we to understand lawhood? What is the relationship between emergent constraints and the constraints imposed by fundamental physical principles? Much work remains to be done, but I hope that the preceding discussion has laid the groundwork.

## Acknowledgements

This work was supported financially by the University of Southern California Dornsife Sustainability Task Force.  Special thanks to Daniel Estrada, Philip Kitcher, David Albert, and Porter Williams for providing patient feedback and discussion of the material under discussion.